\documentclass[aps,prl,preprint,superscriptaddress]{revtex4-1} 
\pdfoutput=1
\usepackage{graphicx}
\usepackage{amsmath}
\usepackage{epstopdf}
\usepackage{hyperref}
\usepackage{color}

\begin{document}

\newcommand{\<}{\langle}
\renewcommand{\>}{\rangle}
\newcommand{\beq}{\begin{equation}}
	\newcommand{\eeq}{\end{equation}}
\newcommand{\kc}{K_{\text{c}}}
\newcommand{\lt}{L_\tau}
\newcommand{\asp}{\alpha_{\tau}}
\newcommand{\br}{{\bf r}}
\newcommand{\bk}{{\bf k}}
\newcommand{\bJ}{{\bf J}}
\newcommand{\bP}{{\bf P}}
\newcommand{\bB}{{\bf B}}

\newcommand{\mw}[1]{{\bf \textcolor{red}{#1}}}

\title{Phase-change switching in 2D via soft interactions}

\author{Rogelio D\'iaz-M\'endez} 
\affiliation{Department of Physics, KTH Royal Institute of Technology,
	SE-106 91 Stockholm, Sweden}

\author{Guido Pupillo} 
\affiliation{icFRC, ISIS (UMR 7006), IPCMS (UMR 7504), Universit\'e de Strasbourg and CNRS, 67000 Strasbourg, France}

\author{Fabio Mezzacapo} 
\affiliation{Univ Lyon, Ens de Lyon, Univ Claude Bernard, CNRS, Laboratoire de Physique, F-69342 Lyon, France}

\author{Mats Wallin} 
\affiliation{Department of Physics, KTH Royal Institute of Technology,
	SE-106 91 Stockholm, Sweden}

\author{Jack Lidmar} 
\affiliation{Department of Physics, KTH Royal Institute of Technology,
	SE-106 91 Stockholm, Sweden}

\author{Egor Babaev} 
\affiliation{Department of Physics, KTH Royal Institute of Technology,
	SE-106 91 Stockholm, Sweden}

\begin{abstract}
We present a new type of phase-change behavior relevant for information storage applications, that can be observed in 2D systems with cluster-forming ability.
The temperature-based control of the ordering in 2D particle systems 
depends on
the existence of a crystal-to-glass transition.
We perform molecular dynamics simulations of models with soft interactions, demonstrating that the crystalline and amorphous structures can be easily tuned by heat pulses.
The physical mechanism responsible for this behavior is a self-assembled  polydispersity, that depends on the cluster-forming ability of the interactions.  
Therefore, the range of real materials that can perform such a transition is very wide in nature, 
ranging
from colloidal suspensions to vortex matter.
The state of the art in soft matter experimental setups, controlling interactions, polydispersity and dimensionality, makes it a very fertile ground for practical applications.    
\end{abstract}

\maketitle


\indent
Ordering processes of particle systems in two dimensions are strongly dependent on the degree of uniformity of the constituents.\cite{lang2010,kelton2010}
For monodisperse ensembles, where all particles are of the same type, crystallization is always reached.
This is the case not only for a slow annealing of the samples but also for temperature-quench protocols, after which the dynamics is characterized by a logarithmic relaxation through the perfect crystal.\cite{kelton2010}
On the other hand, polydisperse ensembles (e.g. particles of different size, charge, etc) favor glass and amorphous states.
Actually, the frustration generated by the polydispersity of the samples can hinder the crystallization even at conditions of quasi-static cooling.\cite{lang2010}
Thus, even if the degree of polydispersity can fill the gap between good crystallizers and good glass-formers, it is impossible to find 2D materials performing well both abilities. 
As a result, 2D systems have not been ideal candidates for phase-change memory applications,\cite{raoux2014} where the fast and reliable switching between amorphous and crystalline states becomes instrumental.\cite{shukla16, salinga2013}

Devices for phase-change memory applications 
are made of bulk chalcogenide alloys, and are 
promising candidates for
both the next generation of high-density ultrafast memories and the emergence of cognitive computing hardware.\cite{koelmans2015} 
The reason for this expectations is the combination of a number of features of phase-change materials, such as high density packing, low power, fast switching, non-volatility and large-scale manufacture.\cite{raoux2014,salinga2013,shukla16,koelmans2015}   
With the application of a proper electrical or heat pulse, a phase-change material can reversibly change from the amorphous to the crystalline phase. 
The structural difference of the configurations implies a change in the resistance or the optical properties, which in turn is used to store information.
As said before, the search for 2D and quasi-2D materials having this properties is still intense, since they could improve many properties and lower the costs.
However, the fundamental problems to overcome, i.e. the structural relaxation of the amorphous phase and the high density of defects in the crystal, are still open problems even for 3D systems.\cite{koelmans2015,ciocchini2016}

Recent studies in 2D monodisperse systems interacting via cluster-forming potentials have demonstrated the existence of a crystal-to-glass transition (CGT) driven by temperature quenches.\cite{dm2017}
After quenches from the disordered state to any final temperature below this transition, an amorphous configuration sets in due to the emergence and preservation of a self-generated polydispersity among the formed clusters.
This phenomenon depends only on the cluster-forming ability of the interactions, e.g. pair-wise interactions displaying a negative minimum in the Fourier transform.\cite{likos2007,glaser2007}
As we will demonstrate here, this CGT behavior implies the coexistence of both good crystallizer and good glass-former regimes in the same system.
With molecular dynamics simulations we show that 
it is easy to switch between such phases
by sudden tunings of the temperature in the form of heat pulses of different energy, which has a direct implication in the field of optical phase-change memories.
The generality of this behavior, that can be found in soft systems of widely different nature, may open a new direction in the research on phase-change memory in two dimensions.




Unlike normal liquid-glass transition, CGT consist of a fast change of the structural arrangement of the system, where the orientational order quickly  goes from long-range to short range, featuring a crystalline-to-amorphous switch. 
However, this transition occurs around a particular value of the degree of frustration of the material (typically its polydispersity), which makes the CGT very unlikely for practical applications.\cite{yunker2010}
For 2D systems, the cluster-forming ability of the potentials has been recently encountered to be responsible of a new type of temperature-driven CGT: melting-quench protocols to a final temperature $T_*$, below the melting point $T_*<T_m$, lead to a fast formation of clusters whose uneven occupation remains constant, acting as an effective polydispersity and hindering the crystallization.\cite{dm2017}
On the contrary, for quenches at moderate temperatures $T_*<T<T_m$, the hopping mechanism is activated and the clusters quickly evolve to a more uniform size distribution where the exchange of particles additionally contributes to the annihilation of topological defects.

\begin{figure}[!tb]
\begin{center}
\includegraphics[width=.7\columnwidth]{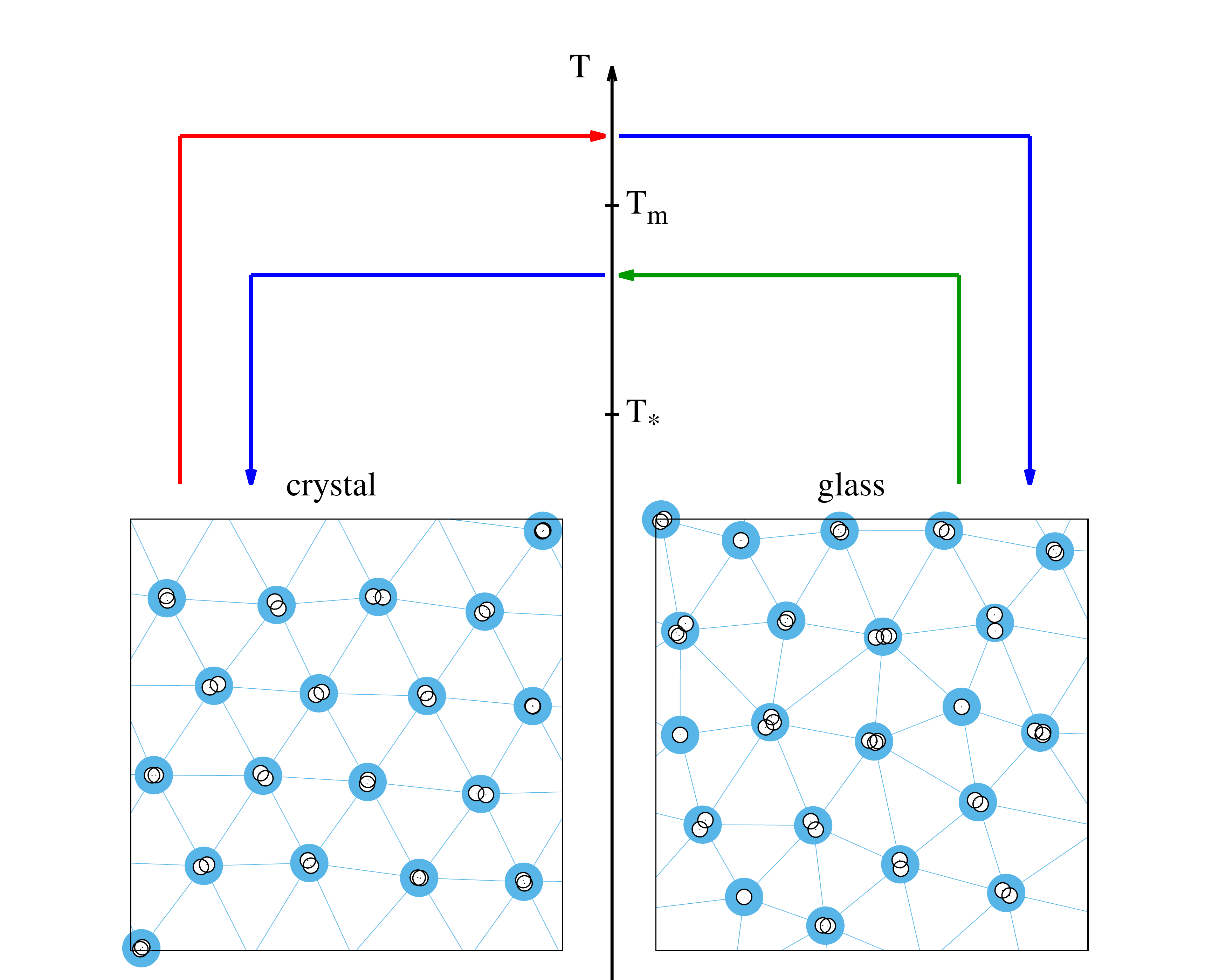}
\caption{
	Snapshots from the simulation of the crystalline phase (bottom left panel) and the glass phase (bottom right panel) at a small temperature $T<T_*$.
	Single particles are represented with void circles and the cluster network is highlighted. 
	The phase switch protocols are indicated with colored arrows:
	from the crystalline to the amorphous phase the system is heated to a high temperature $T>T_m$ in the disordered phase (red arrow) and then rapidly cooled to the initial temperature (blue arrow);  
	from the amorphous to the crystalline phase the system is heated to a moderate temperature $T_*<T<T_m$ in the crystalline phase (green arrow) and then rapidly cooled to the initial temperature (blue arrow).  
}
\label{f1}
\end{center}
\end{figure}

This new phenomenology, in which the structural properties of soft matter systems can be reliably controlled with temperature-based protocols, revives the discussion on the practical use of 2D materials for phase-change applications.   
Figure~\ref{f1} shows the schematic representation of a bit switching in a soft system with cluster-forming ability, operated by heat pulses.
A heat pulse is a process generating a sudden (ideally instantaneous) increase of the temperature up to a value $T_p$, that is further maintained for a    typically short time $\tau_p$ before the initial temperature is restored. 
In practice, phase-change memory applications can use calibrated laser beams or electric currents to control $T_p$ and $\tau_p$.
As depicted in Fig.~\ref{f1}, switching devices operating at glassy temperatures $T<T_*$ can be designed by exploiting the CGT properties of soft matter systems.
A heat pulse with $T_p>T_m$ is indeed equivalent to a melting-quench protocol, provided $\tau_p$ is large enough as to fully disorder the initial configuration whatever it is.
Therefore it is a way of destroying the ordered state, favoring the amorphous configuration.
On the other hand, a heat pulse to a value $T_*<T_p<T_m$ will activate the hopping mechanism and boost the crystallization during the time interval $\tau_p$.
If this time interval is long enough, the crystal structure will eventually emerge as a result of the rearrangement of the clusters, keeping the crystalline phase after the operating temperature is restored.


Clearly, the adequate values of $T_p$ and $\tau_p$ will depend on the particular choice of the system. 
Soft matter models display very dissimilar properties, suggesting the possibility of having a large number of choices for featuring the pulses.
The difference between the structural configuration parameters regarding local and long-range order can also be optimized, as well as the size of the deviations from its average values.
All of these are key features whose control is vital to propose any practical application, and its relations to the specific properties of the interactions have to be extensively explored. 
The quantitative characterization of the local order can be done by looking at the hexatic bond-order parameter of the clusters,\cite{dm2015} which is defined as 
\begin{equation}
\Psi_6=\Big\langle \frac{1}{N_c}\sum_j^{N_c}\frac{1}{N_j}\Big|\sum_l^{N_j}e^{i6\theta_{jl}}\Big|\Big\rangle 
\end{equation}
in analogy to regular (non cluster-forming) crystals. 
Here $N_c$ is the total number of clusters, $N_j$ is the number of clusters neighboring the $j$-th cluster, and  $\theta_{jl}$ is the angle between a reference axis and the segment joining the clusters $j$ and $l$.
As usual, the static structure factor  $S(\mathbf k)=\langle|\sum_{j}^N
e^{i\mathbf{k}\cdot\mathbf{r}_j}|^2/N\rangle$ is calculated through a sum over individual particles and will be used to characterize the long-range orientational order.

As a primary proof of concept,
we focus below on a  
generic interaction potential of the form\cite{dm2015}
\begin{equation}
 U(r)=\frac{U_0}{1+\left(\frac{r}{r_c}\right)^6}
 \label{u}
\end{equation}
This is an ultrasoft potential that approaches the value $U_0$ as the inter-particle distance $r$ decreases below $r_c$, and drops to zero for $r>r_c$,  corresponding  to  soft-core  van  der  Waals  interactions.~\cite{dm2015}
We perform molecular dynamics simulations  
using an overdamped 
Langevin scheme of friction coefficient $\gamma$ 
with up to $N=10000$ monodisperse particles.
Units of length are given in $r_c$, time in $\gamma^{-1}$, temperature in $U_0\times10^{-2}$ and density in $r_c^{-2}$.
At density $\rho=1.4$, the transition temperatures of model (\ref{u}) are $T_m=8$ and
$T_*=4$ 
for the melting and the CGT respectively.~\cite{dm2017}  
It was recently found that similar type of interactions, yielding similar types of phase diagrams
also appears between vortices in  superconductors.~\cite{dm2017} 
In superconductivity it appears in thin films of ``type-1.5" superconductors~\cite{babaev2005semi,babaev2017type}
and in layered systems, where choosing different materials
for different layers should allow to design desired interaction.~\cite{dm2017,varney2013hierarchical,meng2016phase}
The glassy  states  of superconducting vortices, of the type that we discuss
in this paper can be utilized for   cryogenic memory.~\cite{ortlepp2014access,golod2015single}

\begin{figure}[!tb]
\begin{center}
\includegraphics[width=.7\columnwidth]{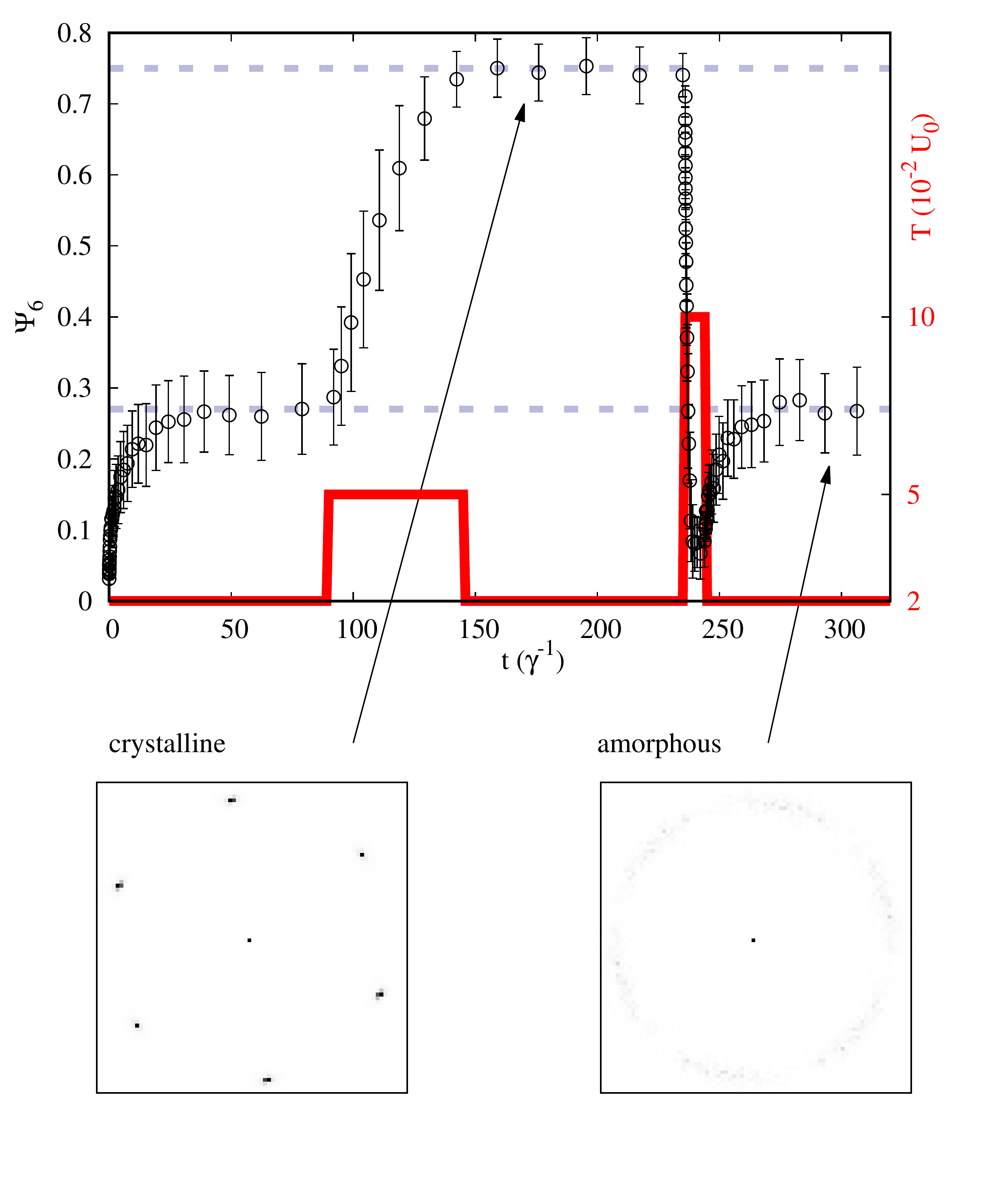}
\caption{Simulation of the bit-writing operation in a system interacting via the potential of Eq.~(\ref{u}).
The local order parameter $\Psi_6$ (black) and the temperature of the thermal bath $T$ (red) are plotted as a function of the time.
The system is prepared at $t=0$ in the amorphous state (bit 0) by an initial quench from a very high temperature to the operating temperature $T=2$.
A first pulse to $T=5$ orders the system in a crystalline configuration (bit 1).
A second pulse to $T=10$ recovers the amorphous state.
Lower panels are typical structure factors in the arrows-pointed configurations.
Dashed lines are guides for the eyes.
}
\label{taus}
\label{f2}
\end{center}
\end{figure}

Figure \ref{f2} shows the simulation of the switching between ordered and disordered phases in the system of Eq.~(\ref{u}), by the application of appropriated heat pulses. 
The temperature of the heat bath (red curve in the figure) is temporarily changed by setting the target value ($T_p$) during certain time-interval ($\tau_p$).
As can be observed from the behavior of $\Psi_6$ (black circles), the designed bit-writing protocols work very well for this model.  
With the proper moderate increase of the temperature the amorphous configuration orders in a crystalline structure, which in turn can be destroyed again by a temperature increase above the melting point.
It is worth noting that the structural change is well defined, i.e. the two configurations are well-separated regarding the quantitative parametrization of both the local and the long-range order.
The differences in the mean values of $\Psi_6$ are well beyond the statistical errors, and the peaks of the structure factor corresponding to the hexatic orientational order of the extended structure drops to less than 5\% when going from the crystalline to the amorphous phase.    

For the model of Eq.~(\ref{u}) at the operating temperature, the crystallization of the amorphous phase is still absent up to times $t>10^5$. 
This is a measure of the quality of the CGT and has to be well established as a primary goal.
For any given model this time is of utmost importance as it is related to the volatility of the recorded information.
The effects of several types and degrees of polydispersity could also be considered to better design stable applications. 
In this case the analysis should be extended to verify the robustness of the crystalline phase.

Not surprisingly, the minimum time required for switching the system (lower-bound for $\tau_p$) is naturally larger when going from the disordered to the ordered configuration than in the  opposite direction.
This is, however, dependent on the target temperature $T_p$ and a case by case study has to be done to evaluate the crystallization speed in the allowed range of temperatures.  
Further, non-instantaneous tuning of the temperature is more realistic and can 
be studied by the methods used here, to
provide better insights into specific switch implementations.

The present paper gives a numerical proof of concept test to demonstrate the feasibility of phase-change applications based on the temperature-driven 
CGT
in soft cluster-forming models.   
Our result enlarge the scope of the search for systems with new phase-change properties in 2D, including a large number of self-assembling systems.
Further numerical simulations and analytical studies of increasingly realistic models will be necessary in order to clearly identify the most promising candidates for experimental verifications.

\section{Acknowledgments}
This work was supported by the Swedish Research Council VR grant 621-2012-3984. 
Computations were performed on resources provided by the Swedish National Infrastructure for Computing (SNIC) at HPC2N.

\bibliography{refpcm}

\end{document}